\documentstyle[aps,prl]{revtex}
\input epsf
\input{epsf.sty}

\begin{document}
\twocolumn[
\hsize\textwidth\columnwidth\hsize
\csname@twocolumnfalse\endcsname
\setcounter{equation}{0}
\hyphenation{pro-vi-ding}

\title{Supercooling of the disordered vortex lattice in 
Bi$_{2}$Sr$_{2}$CaCu$_{2}$O$_{8+\delta}$}
\author{C.J. van der Beek$^{1}$, S. Colson$^{1}$,
M.V. Indenbom$^{1,2}$, and M. Konczykowski$^{1}$}
\address{ $^{{\rm 1}}$Laboratoire des Solides Irradi\'{e}s, Ecole Polytechnique,
91128 Palaiseau, France \\
\noindent $^{{\rm 2}}$Institute of Solid State Physics, 142432
Chernogolovka, Moscow District, Russia}
\date{September 10, 1999}
\maketitle

\begin{abstract}
Time--resolved local induction measurements near to the vortex 
lattice order-disorder transition in optimally doped Bi$_{2}$Sr$_{2}$CaCu$_{2}$O$_{8+\delta}$
single crystals shows that the high--field, disordered phase can be 
quenched to fields as low as half the transition field. Over an 
important range of fields, the electrodynamical behavior of the vortex system 
is governed by the co-existence of the two phases in the sample. We 
interpret the results in terms of supercooling of the high--field 
phase and the possible first order nature of the 
order-disorder transition at the ``second peak''.
\end{abstract}


] 
\narrowtext

It is now well accepted that the mixed state in type II superconductors 
is itself subdivided into different vortex phases. 
In very clean materials, notably single crystals of the  high-$T_{c}$--cuprates 
YBa$_{2}$Cu$_{3}$O$_{7-\delta}$ \cite{Safar92II} and 
Bi$_{2}$Sr$_{2}$CaCu$_{2}$O$_{8+\delta}$ \cite{Pastoriza94,Zeldov95II,Khaykovich96}, 
the vortex lattice, characterized by long-range translational and orientational 
order, undergoes a first order transition (FOT) to a flux liquid 
state without long-range order \cite{Cubitt93}. The FOT is observed at 
inductions $B$ at which vortex pinning by crystalline defects is 
negligible and the vortex system can rapidly relax to thermodynamic 
equilibrium \cite{Zeldov95II,Schilling96}, but which are still much 
below the upper critical field $B_{c2}$. 
The FOT is prolongated into the low temperature regime of 
nonlinear vortex response by a transition from the weakly 
pinned low--field vortex lattice to a strongly pinned, disordered 
high--field vortex          
phase \cite{Khaykovich96,Cubitt94,Nishizaki97}. This 
order-disorder transition is manifest through the so-called ``second peak'' 
feature in the magnetic hysteresis loops, a result of the 
dramatic increase of the sustainable shielding current associated with bulk pinning 
\cite{Khaykovich96,Chikumoto92II,Majer94,Berry97}.  
It was proposed that the crossover from the FOT to the ``second peak'' 
regime constitutes a critical point in the phase diagram \cite{Zeldov95II,Safar93}, 
which in Bi$_{2}$Sr$_{2}$CaCu$_{2}$O$_{8+\delta}$ would lie near $T \approx 40$ K.
In more dirty type-II superconductors
the FOT and the critical point are absent and the critical current ``peak effect'' 
is found at temperatures up to $T_{c}$ \cite{alltogether}. 
The peak effect is often accompanied by strongly history dependent dynamical 
behavior of the vortex system at fields and temperatures just below it, 
suggesting that a first order transition lies at its origin
\cite{Wordenweber86II,Henderson96,Henderson98}.

Among the abovementioned materials the layered superconductor 
Bi$_{2}$Sr$_{2}$CaCu$_{2}$O$_{8+\delta}$ has a specific interest: its high 
Ginzburg number $Gi \sim 0.01$ means that vortex lines are extremely 
sensitive to thermal and static fluctuations and that the FOT and 
second--peak lines are depressed to inductions lower than 1 kG. The 
local induction and flux dynamics around the transition can then be accurately measured 
using local Hall--array magnetometry 
\cite{Zeldov95II,Khaykovich96,Majer94,Berry97} and magneto-optics. Recently, 
the decomposition of the vortex system near the FOT in 
Bi$_{2}$Sr$_{2}$CaCu$_{2}$O$_{8+\delta}$
into coexisting lattice and liquid phases was imaged using this latter technique 
\cite{Soibel99}. In this Letter, we image the flux dynamics 
and the coexistence of the two \em pinned \rm vortex phases 
in Bi$_{2}$Sr$_{2}$CaCu$_{2}$O$_{8+\delta}$ 
near the disordering transition at the ``second peak'' . 
In particular, we find that the disordered phase can be quenched to flux 
densities that are nearly half that at which it exists in equilibrium. 
We interpret our results in terms of supercooling of the high field phase. 
This suggests that the order--disorder transition at the second peak is of first order, 
and that it is the ``true'' continuation of the FOT in the regime of slow vortex dynamics. 
By implication, we propose that a putative critical point lies at a 
temperature not exceeding 14 K.

The experiments were performed on an optimally doped 
Bi$_{2}$Sr$_{2}$CaCu$_{2}$O$_{8+\delta}$
single crystal ($T_{c} = 90$ K) of size $630 \times 250 \times 35$ 
$\mu$m$^{3}$, grown at the University of Tokyo 
using the travelling--solvent floating zone technique, and selected 
for its uniformity. Previous experiments on this crystal
using the Hall-probe array technique 
have revealed the disordering transition of the vortex lattice
to occur at $B_{sp} = 380$ G \cite{Berry97}.
We have visualized the flux density distribution at inductions close to the 
transition using the technique of Ref.~\onlinecite{Dorosinskii92}. A 
ferrimagnetic garnet film with in--plane anisotropy is 
placed directly on top of the crystal, and observed using linearly 
polarized light. The induction component perpendicular to the 
film induces a perpendicular magnetization and concommitant 
Faraday rotation of the polarization, which is then vizualised
using an analyzer. The local induction--variations
induced by the presence of the superconducting crystal are revealed as
intensity variations of the reflected light, the brighter regions 
corresponding to the greater flux, or vortex, density. The technique is 
particularly useful for the study of the low--field behavior of oxide 
superconductors, in which the measurement of the electromagnetic 
response of the vortex system is easily marred by the 
presence of surface barriers \cite{Indenbom94IV,Zeldov94II}, the appearance 
of the Meissner phase \cite{Indenbom95,VlaskoVlasov97}, and macroscopic 
defects \cite{Schuster95III,Koblischka95}. The direct mapping of the flux density 
allows one to distinguish where currents flow inside the 
superconductor\cite{Welp95}, to identify inhomogeneous parts of the crystal, 
and, eventually, to eliminate them.

\begin{figure}
	 \vspace{-10mm}
	\centerline{\epsfxsize 7.5cm \epsfbox{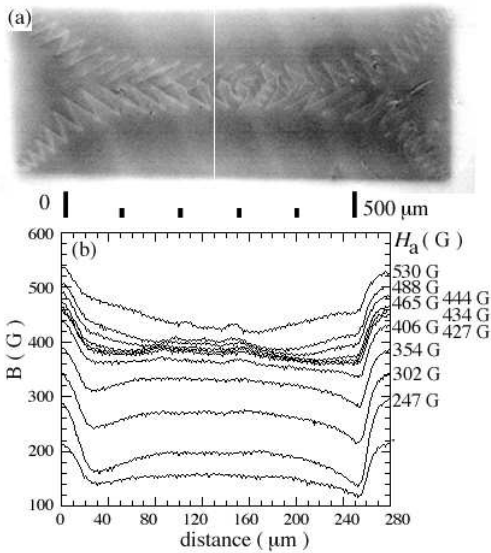}}
	\label{fig:slowramp}
	\caption{(a) Magneto-optical image of the 
	flux distribution on the surface of the 
	Bi$_{2}$Sr$_{2}$CaCu$_{2}$O$_{8}$ crystal after zero--field cooling to $T = 
	24.6 $ K and the slow application of an external field $H_{a} = 
	486$ G. (b) Profiles of the magnetic induction at successive values 
	of the external field during the field ramp, taken along the white 
	line in (a). The step in the 
	induction at either edge of the crystal is the result of the surface barrier
	screening current \protect\cite{Indenbom94IV}. The small 
	irregularities in the center are due to the presence of magnetic domain walls 
	in the garnet film, visible as the ``sawtooth'' structures in (a). \rm}
	\end{figure}

Figure~\ref{fig:slowramp}1(a) shows a magneto-optical image of the 
crystal after zero--field 
cooling to $T = 24.6$ K and the slow ramp of the applied magnetic field 
$H_{a}$ to 486 G. There is a bright belt around the crystal edge, 
corresponding to a region of high field gradient, and, visible under 
the sawtooth--like magnetic domain wall structure in the garnet, an inner region 
with little contrast, indicating a plateau in the local induction. The 
axis of the sawtooth structure is located there where the induction component parallel to 
the garnet film vanishes\cite{Indenbom94III}. This corresponds to the boundary between 
regions of zero and non-zero screening current in the 
Bi$_{2}$Sr$_{2}$CaCu$_{2}$O$_{8}$ crystal. We infer that no  
current flows in the central region of (nearly) constant induction. 
Conversion of the light intensity to flux density shows that the plateau 
induction equals that expected at the transition, $B_{sp} = 380$ G.

The evolution of the flux profiles at successive values of the applied 
field during the ramp is shown in Fig.~\ref{fig:slowramp}1(b). 
At small $H_{a}$, one has a comparatively large step in the induction 
at the crystal edges, and a dome--shaped flux distribution in the 
crystal interior. Such a flux profile occurs when the edge screening 
current due to a surface barrier is much greater than the bulk shielding 
current, which is the result of vortex pinning \cite{Indenbom94IV,Zeldov94II}. 
The dome--like profile moves up to higher induction values as field 
is increased; its evolution stops when the induction in the crystal 
center reaches $B_{sp} = 380$ G (for $H_{a} = 427$ G). As

\noindent    $H_{a}$ is 
increased further, the flux profile flattens out, {\em i.e.} $B$ 
becomes constant throughout the crystal as the high--field vortex 
phase spreads outwards from the crystal center. As a result, the slope 
$\partial B /\partial H_{a}$ becomes equal to the Meissner slope 
\cite{Berry97}. At $H_{a} = 444$ G, the whole crystal is in the 
high--field state, and new flux (vortices) penetrates from the edges; 
it cannot, however, accumulate in 
the crystal center but adopts the linear gradient characteristic of 
the pinning-induced critical state\cite{Bean62}.  This indicates that, at this 
temperature, field, and field ramp rate, the pinning current is comparable to 
or greater than the surface barrier current, giving rise to the ``second peak 
feature'' in the magnetic moment \cite{Khaykovich96,Cubitt94,Berry97}.

\begin{figure}
	\centerline{\epsfxsize 7cm \epsfbox{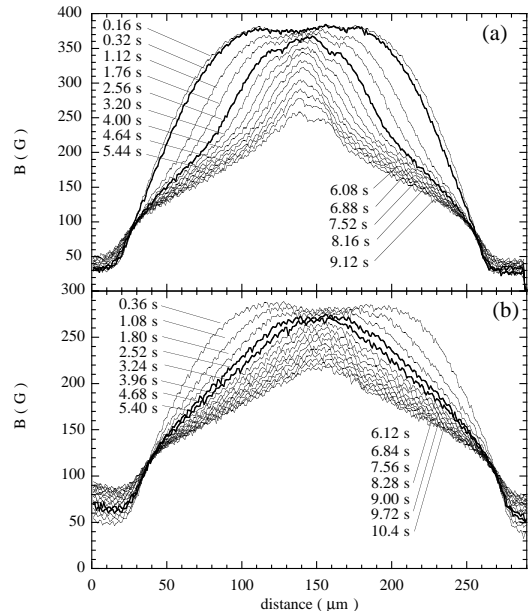}}
	\label{fig:reldown}
	\caption{ (a) Relaxation of the flux profile 
	on the surface of the Bi$_{2}$Sr$_{2}$CaCu$_{2}$O$_{8}$ crystal at $T = 23.0 $ K
	after the application of an external field $H_{a}^{max} = 500$ G and its successive rapid decrease 
	to 120 G. The profiles display three slopes $\partial B_{z}/\partial 
	 x$, corresponding to the surface barrier screening current (near the 
	edges at 30 $\mu$m and 260 $\mu$m)
	\protect\cite{Indenbom94IV}, and 
	to bulk screening currents in the low--field and high--field vortex phases, 
	respectively. (b) The same, but after field--cooling from $T = 
	28$ K in $H_{a}^{max} = 383$ 
	G. The profiles are composed of two linear sections
	corresponding to the surface barrier current and the relaxing critical 
	state established in the low--field phase {\em only}.}
	\end{figure}

Figure~\ref{fig:reldown}2(a) shows the relaxation of the flux profile 
after a rapid decrease of the applied field from 500 G to 
120 G, at $T = 23$ K. The initial flux profile is similar to the one 
for $H_{a} = 488 G$ in Fig.~1: the ``critical state'' fronts 
of the high--field phase have not yet penetrated the whole crystal so
that the induction in the crystal center is nearly constant $B \gtrsim B_{sp}$;
Thus, the internal induction is lower than the applied field because of the 
combined screening by the surface barrier current \cite{Indenbom94IV,Zeldov94II} 
and by the high--field phase. When $H_{a}$ is suddenly decreased,
the sample initially fully screens the field change  ($t = 0.16$ s). From 
$t \gtrsim 0.32$ s onwards, vortices  leave the sample. The flux 

\noindent 
profiles display three distinct linear sections with different 
gradients, corresponding to three mechanisms opposing vortex motion and exit.
The gradient nearest to the crystal edge corresponds to the surface barrier 
current \cite{Indenbom94IV,Zeldov94II}; the two gradients in the bulk correspond 
to the (rapidly decaying) screening current in the low--field vortex lattice phase and 
the (nearly constant) current in the high--field disordered vortex phase, respectively. 
The phase transformation line, at which one passes from the low--field to the high 
field current, progressively moves to the crystal center, until the whole 
crystal is in the low--field phase at $t \gtrsim 10$ s. We note that 
these features are not observed if one prepares a similar initial flux 
profile with a central plateau of $B < B_{sp}$ 
(Fig.~\ref{fig:reldown}2(b)). There are then only \em two \rm flux gradients 
corresponding to the surface barrier and the screening current in the 
low--field phase. These results unambiguously demonstrate that the 
region of constant flux density $B_{sp}$ in the sample center, obtained 
during a slow field ramp (Fig.\ref{fig:slowramp}1), is in the high--field phase; 
namely, it responds to an external field perturbation by developping the 
corresponding, ``high--field'', screening current. Moreover, the 
screening current at any point in the crystal \em depends on the 
history \rm of the vortex system. This is well seen at {\em e.g.} $x = 
100$ $\mu$m and $B = 250$ G: if this induction is attained by a rapid quench from 
the high--field phase, the screening current is equal to that usually observed  
for $B > B_{sp}$. If, during the experiment, the vortex system did not undergo the 
phase transformation to the disordered state, a screening current 
characteristic of the low--field phase is observed.

A glance at the flux density scale on Fig.~\ref{fig:reldown}2 shows 
that the disordered vortex phase in the center of the 
crystal, identifiable in Fig.~\ref{fig:reldown}2(a) by the large screening current it can   
sustain, has in fact been quenched
to inductions nearly half $B_{sp}$.
A similar situation occurs if one rapidly increases the 
magnetic field from zero to a value much

\begin{figure}
	\centerline{\epsfxsize 7cm \epsfbox{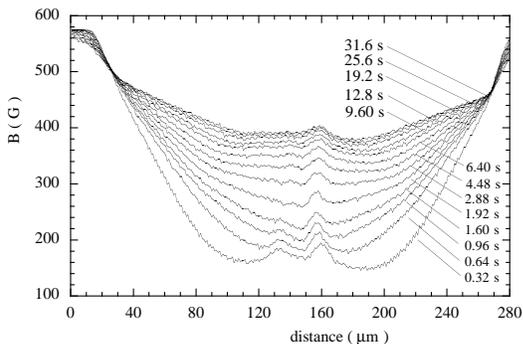}}
	\label{fig:relup}
	\caption{ Relaxation of the induction profile 
	on the surface of the 	Bi$_{2}$Sr$_{2}$CaCu$_{2}$O$_{8}$ at $T = 23.0 $ K
	after the sudden application of an external field $H_{a}^{max} = 537$ 
	G. For $t > 0.96$ s, the flux 
	profiles display three slopes $\partial B_{z}/\partial x$, corresponding 
	to the surface barrier current, and bulk screening currents in the 
	high--field (outer) and low--field (inner) vortex phases, respectively. 
	The irregularities near the center are again due to the presence of domain walls in the 
	garnet film.}
	\end{figure}

\noindent  above $B_{sp}$
(Fig.~\ref{fig:relup}3). Initially, the crystal again perfectly screens the 
field change. As vortices enter the crystal, they are initially in the 
disordered state. Only when they move sufficiently far into the interior does the phase 
transformation to the ordered vortex lattice state take   place. The 
decrease of the flux density from $B \approx \mu_{0}H_{a} > B_{sp}$ 
at the crystal edge to $B \ll B_{sp}$ inside the 
sample due to formation of the critical state necessitates
the presence of the phase transformation line in the sample interior.
This is visible in  Fig. \ref{fig:relup}3 as the changes in the
induction gradient near 80 and 220 $\mu$m. 
The induction at which the transformation takes place 
is again lower than $B_{sp}$, \em i.e. \rm the high field 
phase is now quenched as it penetrates from the sample edge, in this 
case to an induction $\sim 200$ G. As the 
induction gradient in the high field phase relaxes due to thermally 
activated flux motion, the phase transformation line moves from the 
crystal edges towards the crystal centre. In contrast to the dramatic 
quenching of the disordered phase, we did 
not, in any experiment, obtain unambiguous indications that the low field state 
can be prepared at $B > B_{sp}$.

The above observations have important implications for the vortex 
phase diagram in Bi$_{2}$Sr$_{2}$CaCu$_{2}$O$_{8+\delta}$ and 
other layered superconductors. First of all, it is shown that the
phase transition line between the low--field lattice phase and the
high--field disordered vortex phase is, to within our spatial 
resolution ($\sim 10$ $\mu$m), sharp, and that its position can be 
readily identified by the difference in shielding current density 
developped by the two phases after a field pertubation. The phase 
transformation line can, depending on the ratio of these currents,
move inwards from the crystal edge, which happens at 
relatively low temperature or large field sweep rates 
(Figs.~\ref{fig:reldown}2 and 3), 
or outwards from the crystal center, at higher temperatures near the 
reported ``critical point'' \cite{Zeldov95II} or during slow field 
ramps (such as in Fig.~\ref{fig:slowramp}1). In the latter case, any small
``external'' field perturbation is screened by the high current 
developped by the disordered vortex phase, so 
that the induction in the crystal center is held constant and equal 
to $B_{sp}$ (as in Fig.~\ref{fig:slowramp}1). This notably 
holds for the induction change $\Delta B$ associated with the entropy 
change $\Delta S = (1/\Delta B)\partial H_{m}/\partial T$ at the FOT 
\cite{Zeldov95II}($H_{m}$ is the FOT field). Thus, our results give a natural 
explanation for the apparent vanishing of $\Delta S$ at a nonzero 
temperature, without the need to invoke the presence of a critical point in 
the phase diagram. It is not $\Delta S$ that vanishes at $ T \sim 40$ 
K, but the corresponding change $\Delta B$ of the equilibrium flux density 
which cannot be observed, because it is perfectly screened by the pinning 
``critical'' current developped in the disordered high--field phase. 
In other words, thermodynamic equilibrium can no longer be achieved, because, 
for $T \lesssim 40$ K \cite{Fuchs98II}, the high--field phase is pinned on the typical 
experimental time scale. The extra vortices needed to satisfy the 
constitutive relation $B(H)$ cannot enter the region where the 
high--field phase is present.

%
%
Further support for the absence of a critical point near 40 K is given by the 
quenching experiments of Figs.~\ref{fig:reldown}2(a) and 3. 
The flux distributions shown in these plots correspond to
the coexistence of the ordered low--field vortex lattice state and 
the disordered high--field phase. The latter is \em metastable \rm 
since it exists at inductions that at much smaller than $B_{sp}$. We 
interpret this observation as {\em supercooling} of the disordered
state, which in turn suggests that the transition at $B_{sp}$ is of 
first order. Further, the continuity with the high--temperature FOT 
\cite{Khaykovich96} implies that it is simply the continuation of the 
``lattice--to liquid'' transition into the regime of slow vortex 
dynamics. The observation of the present features at 
temperatures down to 14 K, below which the ``second peak'' cannot be 
observed at ordinary experimental timescales, means 
that if a critical point exists, it should lie {\em below} this 
temperature. This would be in agreement with the high--field vortex glass 
transition line of Ref.~\cite{vdBeek92}. The low--field extrapolation of this line, 
was found to interrupt $B_{sp}$ around the same temperature.

%

Finally we point out the consequences for measurements of flux dynamics. 
The possibility of phase coexistence should be taken into account in 
low--field magnetic relaxation experiments, especially those triggered 
by a decrease in the applied magnetic field. In such experiments, the decay rate 
of the global magnetic moment and of the local induction will be determined by no less than 
four contributions: the relaxation of the surface barrier current, 
flux creep in the low--field and high--field vortex phase, and the 
rate at which the vortex lattice recrystallizes at the phase transformation line.
At temperatures below 20 K, these processes become slow and similarly 
impede flux transport. Supercooling of a disordered 
vortex phase has been previously observed in other type-II superconductors such as 
$\alpha$-Nb$_{3}$Ge\cite{Wordenweber86II} and NbSe$_{2}$\cite{Henderson96}. 
The anomalous flux dynamics observed in the field regime close too but 
below the critical current peak \cite{Henderson98} 
may find a natural explanation in the ``asymmetric'' vortex response 
and flux profiles introduced in transport measurements by phase 
coexistence and the supercooling phenomenon.

In conclusion, we have visualized the flux distribution in the ``second 
peak'' regime in Bi$_{2}$Sr$_{2}$CaCu$_{2}$O$_{8}$.
The peak effect feature, the fact that $\partial M /\partial 
H_{a} = -1 $ below the peak, and the vanishing of $\Delta B$ 
associated with the FOT at $ T \sim 40$ K are the result of the pinning current in 
the high field phase, which prohibits flux entry into this phase until 
the phase transformation is complete. We have observed supercooling of 
the high--field disordered vortex system to fields nearly half the 
phase transformation field $B_{sp}$. The results suggest that the 
vortex order--disorder transition at the second peak in  Bi$_{2}$Sr$_{2}$CaCu$_{2}$O$_{8}$ 
is first order, and that any critical point in the phase diagram lies 
below 14 K.



We thank N. Motohira for providing the \linebreak 
Bi$_{2}$Sr$_{2}$CaCu$_{2}$O$_{8}$ crystal, and M.V. Feigel'man, 
P.H. Kes, A. Soibel, A. Sudb\o, and E. Zeldov for fruitful discussions.

\vspace{-1mm}

	\end{document}